\documentstyle[prl,aps]{revtex}

\def\beq{\begin{equation}}
\def\eeq{\end{equation}}

\def\reff#1{(\ref{#1})}

\def\vekt#1{\bbox{#1}}

\def\vektr{\vekt{r}}

\def\halb{\frac{1}{2}}

\def\Edach{\hat{E}}
\def\Ptildeplus{\tilde{P}^{+}}
\def\tlaseron{t_{\mbox{\rm\scriptsize on}}}
\def\Ecrit{E_{\mbox{\rm\scriptsize crit}}}
\def\Zeff{Z_{\mbox{\rm\scriptsize eff}}}
\def\epsiloneff{\epsilon_{\mbox{\rm\scriptsize eff}}}
\def\Pseq{P_{\mbox{\rm\scriptsize sequ}}}
\def\Pks{P_{\mbox{\rm\scriptsize KS}}}
\def\Pcl{P_{\mbox{\rm\scriptsize cl}}}
\def\Ptildecl{\tilde{P}_{\mbox{\rm\scriptsize cl}}}
\def\Pkstot{P_{\mbox{\rm\scriptsize KS}}^{\mbox{\rm\scriptsize tot}}}

\def\energy{{\cal{E}}}

\def\ksenergy{{\cal{E}}^0_{\mbox{\rm\scriptsize KS}}}

\def\pabl#1#2{\frac{\partial #1}{\partial #2}}

\def\atomu{\mbox{\rm \ a.u.\ }}

\begin{document}
\draft
\title{A two-dimensional, two-electron model atom in a laser pulse: \\ exact
treatment, single active electron-analysis,\\ time-dependent density functional
theory, classical calculations,\\ and non-sequential ionization}
\author{D.~Bauer}
\address{Theoretical Quantum Electronics (TQE)\cite{www}, Technische Hochschule
Darmstadt,\\ Hochschulstr.\ 4A, D-64289 Darmstadt, Germany}
\date{\underline{\tt Published in Phys.\ Rev.\ A {\bf 56}, 3028 (1997) }}

\maketitle

\begin{abstract}
Owing to its numerical simplicity, a two-dimensional two-electron model atom, with each electron moving
in one direction, is 
an ideal system to study non-perturbatively a
fully correlated atom exposed to a laser field. 
Frequently made assumptions,
such as the ``single active electron''-approach and calculational
approximations, e.g.\ time dependent density functional theory 
or
(semi-) classical techniques, can
be tested. In this paper we examine the multiphoton short pulse-regime.
We observe ``non-sequential'' ionization, i.e.\ double
ionization at lower field strengths as expected from a sequential,
single active electron-point of view.  
Since we find non-sequential ionization also in purely classical simulations,
we are able to clarify the mechanism behind this effect in terms of
single particle trajectories. 
\end{abstract}


\section{Introduction}
Several theoretical approaches were able to reproduce experimentally
observed ion yields in multi-electron ionization,  at least
qualitatively (see e.g.\ \cite{augst}). Most of them are based on a
``single active electron'' (SAE) point of view \cite{schafer,yang}. 
A new impact on the research in this field had the
discovery of the so-called ``knee'' or ``shoulder'' in the ionization
yields of helium exposed to a laser pulse \cite{fitting}. This means that
double ionization occurs {\em many orders of magnitude} more strongly at intensities where,
according to a sequential SAE scenario, almost no
He$^{++}$ should be present. Early after the experimental observation
of this non-sequential ionization (NSI), two possible mechanisms were suggested in order
to explain it. Corkum proposed a
rescattering scenario \cite{corkum} where the first electron revisits the
core and ionizes the second electron collisionally. Fittinghoff {\em
et.\ al.}\ suggested a ``shake off'' effect \cite{fitting} where the second
electron ionizes due to the sudden loss of screening of the core by the first
electron. Walker {\em et.\ al.}\ \cite{walker} concluded by analyzing their
experimental data that a rescattering process is not able to explain
the observed yields. Their arguments are based on the absence of a
rigorous threshold in the He$^{++}$ yields.  Instead they propose
``that NSI occurs via a simultaneous two-electron ejection either
through a shake off or threshold mechanism involving some form of
electron correlation''. Recently, the NSI mechanism has been clarified
within the intense-field many-body $S$-matrix theory \cite{beckerfaisal}. It was
shown ``that the dominant mechanism behind the observed large
probability of laser-induced double escape is a quantum mechanical
process of absorption of photon energy by one of the electrons which
is shared cooperatively with the other electron through the Coulomb
{\em correlation}''. This mechanism for the NSI process was
independently deduced from 1D He-studies where the model atom had
been exposed to a low frequency, short pulse laser field \cite{lappas}:
``[...] before the outer electron disappears completely, the inner
electron is already sufficiently strongly excited so that it leaves
the atom within a short time interval later. It is during this time
interval that the correlated double ionization takes place.'' 
Simulations where the outer electron is calculated in the SAE way but
the inner one feels (in a second computer run) the time dependent
potential created by the outer one, succeeded in reproducing the
NSI-``knee'' \cite{icompvii}. This result is also a strong indication that the
suggested mechanisms as quoted above are, indeed, the correct ones.  
However, there is no detailed physical picture how this energy sharing
between the outer and the inner electron takes place.

Our calculations were performed for a  relatively
high frequency ($\omega=0.4$~a.u.) and a very short pulse duration (6
optical cycles) while in \cite{lappas} a low frequency short pulse was used. 
Since we are rather in the multiphoton-regime than
in the tunneling domain, the occurrence of NSI might be
surprising at all. Indeed, in our calculations NSI is relatively weak
compared to the many orders of magnitude effect for ionization
of helium in strong low frequency laser light.
However, with the help of our additional classical
simulations we are able to provide (i) a detailed physical picture how
NSI takes place in terms of one-particle trajectories, and
(ii) a proof that NSI, in its essence, is not a quantum mechanical effect.

Because the full quantum mechanical numerical simulation of helium
exposed to a laser field is an extremely demanding task \cite{parker}, 
approximate approaches are desirable. Among these,
Hartree-Fock-
\cite{kulander_i,kulander_ii,pindzola_i,pindzola_ii,pindzola_iii},
time-dependent density functional- (TDFT) \cite{gross_i,erhard,ullrich_i,ullrich_ii} and
semi-classical molecular dynamics-calculations \cite{wasson,lerner_i,lerner_ii,lerner_iii} are most
frequently used. Especially the latter method succeeded in reproducing
the ``knee'' \cite{lerner_i}.  
On one hand the molecular dynamics calculations
are very appealing and instructive since particle trajectories and
the single particle energies can be traced. On the other hand the
additional ``Heisenberg-force'' which must be introduced in order to
avoid instabilities where one electron falls into the ``black hole'' (i.e.\ the
nucleus) while the other one ionizes, is somewhat artificial and may
evoke objections against the results produced by this method.

The time-dependent Hartree-Fock method was found to be problematic in
the framework of multi photon ionization \cite{pindzola_i,pindzola_ii,pindzola_iii}. 
Results from TDFT, in
principal an exact approach, depend on the choice of the
effective exchange-correlation potential \cite{ullrich_iii}. Another disadvantage
of this procedure is that  only the total electron density
$n(\vektr,t)=\sum_i \vert\varphi_i(\vektr,t)\vert^2$ is calculated
and the single particle orbitals $\varphi_i(\vektr,t)$ are physically
meaningless in a rigorous sense.

The study of systems where the motion of each electron is reduced to
one spatial dimension has a relatively long tradition. Potentials of
the form $-Z/\sqrt{x^2+\epsilon}$, so-called ``soft core
Coulomb potentials'',  provide an energetic Rydberg-like scaling \cite{javanainen}
and lead to results, qualitatively similar to those from full 3D calculations. 
Two 1D electrons are a two-dimensional system which is tractable with
nowadays computers. Two 1D electron-systems are used to study
non-perturbatively autoionization \cite{schultz}, ionization of a negative ion
\cite{grobe}, validity of time-dependent Hartree-Fock theory for the
multiphoton ionization of atoms \cite{pindzola_i,pindzola_ii,pindzola_iii}, and, most recently,
two-electron effects in harmonic generation and ionization \cite{lappas}.  

This paper is organized as follows.  In Section \ref{modelintro} the model
system is introduced. In Section \ref{qmresults} results from the 2D quantum
calculations are presented. Section \ref{saeanalysis} is devoted to an analysis of
the results in terms of a SAE-approach. In Section
\ref{tddft} we present the results from a time-dependent density
functional-calculation and in Section \ref{classical} we discuss our classical
particle simulations within which the NSI scenario can be clarified.
Finally, we summarize and conclude in Section \ref{concl}.

\section{The 1D helium model} \label{modelintro}
The two 1D electrons with coordinates $x$ and $y$ interact with the core and with each other
through a ``soft core''-interaction, i.e.\ $-2/\sqrt{x^2+\epsilon}$
and $1/\sqrt{(x-y)^2+\epsilon}$, respectively, and with the field
$E(t)$ through the dipole term $(x+y)E(t)$ (atomic units (a.u.) will
be used throughout this paper). 
Thus the total Hamiltonian reads 
\beq
H(x,y,t)=-\halb\pabl{^2}{x^2}-\halb\pabl{^2}{y^2}-\frac{2}{\sqrt{x^2+\epsilon}}-\frac{2}{\sqrt{y^2+\epsilon}}+\frac{1}{\sqrt{(x-y)^2+\epsilon}}+(x+y)E(t).
\label{hamiltonian} \eeq
The desired ground state energy can be tuned by varying $\epsilon$. We
used 
\[ \epsilon=0.55 \] 
in our calculations which leads to the ground state energy
\[ \energy_0=-2.897\atomu \]
on our numerical grid. $\energy_0$ is approximately the ground state
energy for the real 3D helium atom which is $-2.902$.

One may prefer thinking  in
terms of {\em one 2D particle} which moves in the somewhat peculiar 2D
potential
\beq
V(x,y,t)=-\frac{2}{\sqrt{x^2+\epsilon}}-\frac{2}{\sqrt{y^2+\epsilon}}+\frac{1}{\sqrt{(x-y)^2+\epsilon}}+(x+y)E(t)
\label{2dpot}
\eeq
instead of the two electrons interacting with each other.
The potential \reff{2dpot}  and the ground state energy level are shown in Figure
\ref{potentials} for the three constant fields $E=0.0$, $0.1$, and $0.616$. The electric field
$E$ tilts the fieldfree potential around the axis $y=-x$.

In order to estimate at what fieldstrengths strong single and double
ionization should occur, it is advantageous to calculate the classical
critical fields. However, the commonly used method of equating the initial
ground state energy level to the maximum of the barrier formed by the atomic
potential and the external field leads to an unphysically small
critical field $\Ecrit^+=0.009$ for single ionization. In our
numerical simulations in the next sections we will find a negligible
ionization probability for such low field strengths. Like in the
case of hydrogen-like ions, the electron is not able to move beyond the
point where the energy level touches the lowest point of the energetic
barrier \cite{bauer}. Therefore the ``real'' critical field is higher. 
On the other hand, total ionization should be strong at latest when
the ground state energy level exceeds the electron-electron repulsion
ridge in $x=y$-direction. This is the case at $E^{++}=0.616$ (see Figure \ref{potentials}c).

\section{Quantum calculation} \label{qmresults}
We used a spectral method \cite{feit} to solve the time-dependent
Schr\"odinger equation (TDSE) 
\[ i\pabl{}{t}\Psi(x,y,t)=H(x,y,t)\Psi(x,y,t) \]
with the Hamiltonian \reff{hamiltonian}. In order to keep the
numerical effort as small as possible we chose a rather high laser
frequency
\[ \omega=0.4 \]
and a very short pulse covering 6 cycles,
\[ T=6\frac{2\pi}{\omega}=94.248\ . \]
The pulse envelope was $\sin^2$-shaped, thus
\beq E(t)=\Edach \sin^2(\frac{\pi}{T}t)\sin\omega t  \label{efield} \eeq
for $0<t<T$.
With the frequency chosen and $\Edach$ not greater than 1 we are in
the multiphoton-domain since the Keldysh-parameter
$\sqrt{\vert\energy_0\vert/(2 U_p)}$ is not much less than unity over
the whole intensity region of interest. $U_p$ is the ponderomotive
potential $\Edach^2/(4\omega^2)$, i.e.\ the mean quiver energy of a
free electron in the laser field.
A spatial grid spacing $\Delta x= \Delta y = 0.4$ and a temporal one
$\Delta t=0.1$ was found to be sufficient. We chose the grid size large
enough so
that almost no probability density reaches the boundaries within the
pulse time $T$. 
The ground state wave function was determined by propagating a
Gaussian seed function in imaginary time. A contour plot of the ground
state probability density is shown in Figure \ref{groundstate}. Its energy is $\energy_0=-2.897$.

If we assume that one electron is already ionized we are left with
the 1D version of hydrogen-like helium He$^+$. The ground state energy
of this system,
\beq i\pabl{}{t} \Psi^+(x,t)=\left( -\halb \pabl{^2}{x^2}
-\frac{2}{\sqrt{x^2+\epsilon}} \right) \Psi^+(x,t),
\label{schr_heplus} \eeq
 was determined to be $\energy^+=-1.920$. Therefore, the ionization
 energy for the ``outer'' electron is $\energy_0-\energy^+=-0.977$. The
 removal energy for the outer electron in real 3D He is 0.9~a.u.

The probability density $\vert\Psi(x,y,t)\vert^2$ during the pulse for
the peak field
strength $\Edach=0.3$ is shown in  Figure \ref{timeseries}.
One looks perpendicularly from above onto the illuminated,
logarithmically scaled probability density.

Density flowed along the $\pm x$- or $\pm y$-axis corresponds to
single ionization while double ionization happens when both $\vert
x\vert$ and $\vert y\vert$ are significantly greater than the width of
the ground state. 
We use this simple picture to define our ionization
probabilities at the end of  the laser pulse for total, double and single ionization $P$, $P^{++}$
and $P^+$, respectively,
\beq P=1-\int_{-a}^a\!\int_{-a}^a \vert\Psi(x,y,T)\vert^2\, dx\, dy,
\label{totalioni} \eeq
\begin{eqnarray} P^{++} &= 1 & -\int_{-\infty}^{\infty}\!\int_{-a}^a
\vert\Psi(x,y,T)\vert^2\, dx\, dy\nonumber \\ 
&& -\int_{-a}^a\!\int_{-\infty}^{\infty} \vert\Psi(x,y,T)\vert^2\,
dx\, dy\nonumber \\ 
&& +\int_{-a}^a\!\int_{-a}^a \vert\Psi(x,y,T)\vert^2\, dx\, dy,
\label{doubleioni}
\end{eqnarray}
\beq P^{+}=P-P^{++}. \label{singleioni} \eeq
We chose $a=5$. In Figure \ref{integrationareas} the integration areas corresponding to  $P$, $P^{++}$
and $P^+$ are shown.

Of particular interest is the evolution of the probability density when the double
ionization regions $\vert x\vert>5$ {\em and} $\vert y\vert >5$ are
occupied for the first time. In Figures
\ref{timeseries_detail} and \ref{timeseries_detail_0.7} these time
intervals are shown for $\Edach=0.2$ and $0.7$, respectively. The delay between
the snapshots is a quarter optical cycle in each case.  

Let us first analyze the $\Edach=0.2$-case in Figure
\ref{timeseries_detail}. Till $n=2.75$ optical cycles mainly single
ionization has occured, i.e.\ the probability density is still located
along the axes. At time $n=3.00$ two density jets enter the region
$x,y>0$ (indicated by two arrows). They are clearly separated by the
Coulomb repulsion ridge along $y=x$. These density jets
represent states where both electrons are on the same side of the
nucleus. Therefore the Coulomb repulsion is relatively high and the
jets tend to flow back towards the single ionization channels.
This reflux of probability density is supported by the electric field which
has its maximum a quarter of a cycle later, at
$n=3.25$. However, some density passes the single ionization 
channels and appears in the regions $x>0, y<0$ and $x<0, y>0$,
respectively (see arrows in the $n=3.25$- and in the $n=3.5$-plot). 
Now, the two electrons
represented by this density are on opposite sides of the nucleus, and
both are ionized. So we conclude that although the electric field
amplitude $\Edach$ is not sufficiently strong to double ionize by simply
tilting the $xy$-plane, Coulomb repulsion {\em and} electric field can
lead to double ionization by acting together constructively.   
Therefore we propose {\em Coulomb repulsion assisted laser
acceleration} to be responsible for NSI.

At higher intensities ($\Edach=0.7$, Figure
\ref{timeseries_detail_0.7}) we also observe the two density jets entering regions of
the $xy$-plane where $x$ and $y$ have equal signs (see $n=2.00$ till
$n=2.25$). However, reflux of these jets is obviously not essential to
stimulate subsequent emission of probability density into regions
$x>0, y<0$ and $x<0, y>0$, respectively. This can be seen in plot
$n=2.25$ where already bursts of density leave the single ionization
channels although the jets are not flown back yet. Here, double
ionization is mainly due to the strongly tilted $xy$-plane.

Note that these backflowing jets which support double ionization are
of course absent in any SAE model while they are expected to be included in
TDDFT since there both electrons are allowed to respond to the field
simultaneously. Furthermore, purely classical simulations should show
a similar NSI scenario as the quantum density current in the
$xy$-plane does, i.e., electrons which collide while moving in the
same direction (the two jets) followed by  a subsequent turn of one
electron which crosses the origin and finally leaves in the opposite
direction.
All these presumptions  will be confirmed in the next sections.

We performed several runs varying the peak fieldstrength $\Edach$. In
Figure \ref{ioni_yields_I} the ionization degrees $P$, $P^{+}$, and $P^{++}$ are
plotted vs.\ the intensity $I=\Edach^2$. There are also shown the
ionization probabilities $\Pseq$ and $P\cdot\Pseq$ which results from solving
\reff{schr_heplus} in the external laser field. As long as $P^{++}$ is small $P\cdot\Pseq$ would
be the probability for He$^{++}$ production if double
ionization occurs purely sequential, i.e., the first electron does not
interact with the residual ion anymore and the second electron remains
non-excited after the emission of the first one. Obviously this
completely sequential scenario cannot explain the correct ionization
degree $P^{++}$. Double ionization occurs earlier than predicted by
$P\cdot\Pseq$. This corresponds to the experimentally observed
``knee'' in the He ion yields, although our He$^{++}$-curve 
does not show such a pronounced shape due to the fact that we are
in the multiphoton and not in the tunneling regime as in
\cite{lappas}. We have also performed some runs using $\omega=0.2$ and
saw an increasing deviation from the SAE He$^{++}$ curves, i.e.\ an
increasing ``knee''.
The $\Pseq$ and $ P\cdot\Pseq$ curves cross the correct $P^{++}$
result since depletion of the He$^+$ ions is not taken into account.

\section{Single active electron analysis} \label{saeanalysis}
In the previous section we have already compared the correct
ionization degree $P^{++}$ with the product $P\cdot\Pseq$ for the rate
when the second
electron ionizes from its He$^+$ ground state without
interacting with the electron already freed. The ionization degree
$P\cdot\Pseq$ was found to be too small in the intensity region of
interest $0.02<I<0.07$, as depicted in Figure \ref{ioni_yields_I}.

Now, we may try to describe the ionization probability $P^{+}$ in the
single active electron approach. In order to do this we have to find
an appropriate $\Zeff$ and to solve the equation
\beq  i\pabl{}{t} \Psi^0(x,t)=\left( -\halb \pabl{^2}{x^2}
-\frac{\Zeff}{\sqrt{x^2+\epsilon}} \right) \Psi^0(x,t).
\label{schr_henull} \eeq
The first electron has an ionization potential
$\energy_0-\energy^+=-0.977$. 
We found that $\Zeff=1.117$ yields such a
binding energy. 
In Figure \ref{ioni_yields_II} the resulting ion yields $\Ptildeplus$ are shown and compared with
the exact result $P^{+}$. The ionization degree calculated from
\reff{schr_henull} is too high over the whole region where mainly
single ionization takes place. As soon as double ionization occurs the
curve differs also qualitatively from the exact one: there is a dip in the exact
$P$ and $P^{+}$ curves around $I=0.03$ which is absent in the
$\Ptildeplus$ result. We conclude that the second electron shares some
energy with the escaping first electron which leads to a decreased
single ionization probability. Since in a similar work for a very low
frequency \cite{lappas} the SAE curves are found to fit well, the deviations
we observe might be mainly due to the relatively high frequency we have
chosen. For higher frequency laser fields  the ``cracking
up'' of the initial ground state wave function into an ``inner'' and
an ``outer'' electron may be too slow to be
well described by an SAE ansatz for the ``outer'' electron.

One may object that although tuning   $\Zeff$ in order to fit the binding
energy of the ``outer'' electron leads to an over-estimation of the
ion yields, there might be a certain combination of $\Zeff$ and an
effective soft-Coulomb-parameter $\epsiloneff$ which does both:
providing the correct binding energy {\em and} the right ionization
probabilities. However, we tried $\Zeff=1$ and $\epsiloneff=0.398$ which leads
also to the desired binding energy $-0.977$. The resulting curve for
the single ionization yields are also shown in Figure \ref{ioni_yields_II}. It
overestimates the single ionization too.

\section{Density Functional Theory} \label{tddft}
Density functional theory (DFT) is a powerful tool in determing
multielectron atomic structures (see e.g.\ \cite{gross_i,dreizler} for an
overview). It has been shown \cite{runge} that a Hohenberg-Kohn-type
theorem exists also for time-dependent phenomena. Therefore the
existence of an effective potential which transforms the problem of
$N$ interacting electrons to that of $N$ non-interacting ones is
proved.
The non-interacting electrons move in an effective potential which is
a functional of the total electron density only. The problem reduces to
finding an appropriate effective potential which includes all relevant
exchange and correlation effects. 

There is some recent work on the field of time-dependent density
functional theory (TDDFT) applied to laser ionization of atoms \cite{gross_i,erhard,ullrich_i,ullrich_ii}.   

In the case of a singlet two-electron-system like our model atom of helium, there is only one occupied Kohn-Sham-orbital $\varphi(x,t)$. The
total electron density is
\beq n(x,t)=2\vert \varphi(x,t)\vert^2. \label{dens_orb} \eeq
There are no exchange contributions, and neglecting correlation
effects leads to the time-dependent Hartree equation
\beq i\pabl{}{t} \varphi(x,t)=\left( - \halb \pabl{^2}{x^2} - \frac{2}{\sqrt{x^2+\epsilon}}+ \int 
\frac{\vert \varphi(x',t)\vert^2}{\sqrt{(x-x')^2+\epsilon}}\, dx' + xE(t)
\right) \varphi(x,t). \label{kohnsham} \eeq

Hartree's independent electron-model was used by Geltman already in 1985 to analyze
experimental results in the multiple ionization of atoms \cite{geltman_i,geltman_ii}.

The ground state energy we obtained by solving \reff{kohnsham} in
imaginary time is $\ksenergy=-2.878$.

In Figure \ref{tddftcomp} the comparison between the TDDFT-results and the exact
ones are made. The ionization yields for He$^+$ and He$^{++}$ are
observables, of course. Thus they are functionals of the density
$n(x,t)$ and, due to the simple relation  \reff{dens_orb}, explicit
functionals of the Kohn-Sham-orbital density
$\vert\varphi(x,t)\vert^2$. We want do adopt the simple
``integration-over-areas'' picture in order to calculate the
ionization, as described in Section \ref{qmresults}. According to Ullrich {\em et.\ al.}\
\cite{ullrich_i} we proceed, for the time being,  as follows: with 
\beq \Pks:=1-\int_{-a}^a \vert \varphi(x,t)\vert^2\, dx
\label{ks_integration} \eeq  
the probability for neutral helium is the product of the probabilities
for each orbital to be  non-ionized. Thus
\beq \Pks^0=(1-\Pks)^2. \label{ullr_neutral} \eeq
For the single and double ionization 
\beq \Pks^+=2 \Pks (1-\Pks), \label{ullr_single} \eeq
\beq \Pks^{++}=\Pks^2 \label{ullr_double} \eeq
follows and the total ionization clearly is
\beq \Pkstot=1-\Pks^0. \label{ullr_total} \eeq 
The three curves corresponding to $\Pkstot$, $\Pks^+$ and
$\Pks^{++}$ are shown in Figure \ref{tddftcomp}. The agreement is quite bad. The
total and single ionization degree are overestimated as in the
SAE calculation.

However, the total ionization $\Pks$ fits well the exact $P$ if
one avoids assigning any physical relevance to the Kohn-Sham-orbital
$\varphi(x,t)$ and proceeds instead as follows: $\Pks$ as defined in
\reff{ks_integration} is the probability to find {\em any} of the two
electrons outside the interval $[-a,a]$ since the {\em physical}
total electron density is
$n(x,t)=2\vert\varphi(x,t)\vert^2$. Therefore the probability for
total ionization should be simply
\beq \Pkstot=\Pks, \label{my_total} \eeq
and 
\[ \Pks^0=1-\Pks . \]
$\Pkstot$ according \reff{my_total} is also depicted in
Figure \ref{tddftcomp}. The agreement between $\Pks$ and $P$ is
excellent. 
The dip around $I=0.03$ is
well reproduced. Since the dip is absent in the SAE results, the onset
of NSI seems to be included in the TDDFT although only the simple
Hartree effective potential is taken. 

There is no simple way to deduce $\Pks^+$ and
$\Pks^{++}$ without claiming physical significance of the
Kohn-Sham-orbital $\varphi(x,t)$ although the {\em existence} of purely
density dependent functionals $\Pks^+[n]$ and $\Pks^{++}[n]$ are
proved by the Hohenberg-Kohn-theorem. Eqs.\
(\ref{ullr_neutral}--\ref{ullr_total}) would be valid if the correct
wave function $\Psi(x,y,t)$ was the product of the Kohn-Sham-orbitals,
$\varphi(x,t)\varphi(y,t)$. A plot of $\vert
\varphi(x,T)\vert^2\vert
\varphi(y,T)\vert^2$ for $\Edach=0.3$ is shown in Figure \ref{dftprobab_i}. Clearly,
there is a gridlike pattern imprinted due to the construction of
the wavefunction as a pure product. This leads to a totally different
angular distribution in the $xy$-plane. A similar behaviour is
observed in time-dependent unrestricted Hartree-Fock calculations
\cite{pindzola_iii}.

However, for higher field strengths both electrons are ionized rapidly
and subsequently behave as free and almost independent electrons. Thus
the total wave function should develop a grid-like pattern. This is
shown in Figure \ref{highfieldprobabs} where the peak field strength $\Edach=0.7$ was
chosen. The TDDFT result is also plotted.

\section{Classical Simulations} \label{classical}
We solved the classical equations of motion according the Hamiltonian
\reff{hamiltonian} and the electric field \reff{efield} for a
microcanonical ensemble of the two electrons. We traced the
one-particle-energies
\begin{eqnarray}
\energy_x(x,y) &=& \halb \dot{x}^2 -\frac{2}{\sqrt{x^2+\epsilon}}+
\frac{1}{\sqrt{(x-y)^2+\epsilon}}, \label{single_energy_x} \\
\energy_y(x,y) &=& \halb \dot{y}^2 -\frac{2}{\sqrt{y^2+\epsilon}}+
\frac{1}{\sqrt{(x-y)^2+\epsilon}}.\label{single_energy_y}
\end{eqnarray}
The total energy is
\[
\energy(x,y,t)=\energy_x(x,y)+\energy_y(x,y)-\frac{1}{\sqrt{(x-y)^2+\epsilon}}+(x+y)E(t).
\]
Each electron is considered to be  ionized when $\energy_x(x,y)>0$ or
$\energy_y(x,y)>0$, respectively. There is no unique way in defining
the single particle energies
(\ref{single_energy_x}--\ref{single_energy_y}) since the
electron-electron term may be shared between the two electrons in
various ways \cite{wasson}. However, this has little influence on the
ionization degrees since in the case of single
or double ionization the distance between
the two electrons is normally large at the end of the pulse.

The initial conditions were chosen to meet the quantum mechanical
ground state energy $\energy_0=-2.897$.
Fortunately, the resulting ion yields were not sensitive to the choice
of the ensemble of initial conditions. Instead of taking several
initial positions and momenta we started with one
``mother''-configuration at $t=0$ and vary the time $\tlaseron$ where the laser pulse
sets in. We tried several mother-configurations to ensure
insensibility of the resulting ion yields.

We would like to mention that a classical treatment of an 1D model helium
has also been undertaken in \cite{lewenrza} in the framework of
stabilization of multielectron atoms.

The results are shown in Figure \ref{classioni}. The single ionization is
strongly overestimated. This is due to the fact that the ionizing
electron gains energy at the expense of the still bound electron which
occupies a state of quantum mechanically forbidden low energy, i.e.\
an energy below $-1.920$. This is
shown in Figure \ref{classsingle} where the two one-particle-energies $\energy_x$
and $\energy_y$ are plotted
for a representative single ionization event at $\Edach=0.1$.

This behavior could be prevented if one introduces a velocity
dependent ``Heisenberg''-potential \cite{wasson}. However, for our purpose to
study the NSI-mechanism this is not necessary. We have calculated
also the classical SAE single ionization process in the potential
$V(x)=-\Zeff/\sqrt{x^2+\epsilon}$ with $\Zeff=1.117$ as in Section
\ref{saeanalysis}. The resulting yields are lower and show a rapid increase at the
critical intensity which is $0.05$ for the potential used. This is the
typical behavior in pure classical simulations. Note that
double ionization is already observed where according the SAE approach
even no classical single ionization occurs. 

The result of the SAE calculation He$^+\to$He$^{++}$ was multiplied with
the probability for He$^+$-production from the full 2D run. Below
$I=0.25$, the classical critical intensity for sequential
He$^{++}$-production the probability vanishes, as expected. 
Thus the intensity region $0.03<I<0.2$ is the classical NSI regime we are
particularly interested in. We examined each double ionization event
in that region. The ionization dynamics of two representative
examples in terms of single-particle-energies and trajectories is depicted in Figures
\ref{classicalnsi} and \ref{classicalnsi_traj}.

For the purpose of comparing our classical results with the quantum
mechanical probability density current in the $xy$-plane we look at
the electron trajectories in Figure \ref{classicalnsi_traj}. 
One observes that the electron
which leaves first is accompanied by the other electron moving  in
the same direction (corresponding to density flowing into regions of the $xy$-plane
where $x$ and $y$ have equal sign). 
The Coulomb interaction is strong within that
half cycle. Then, as the first electron ionizes, the second one turns
around and moves in the opposite direction (corresponding to
probability density passing one of the $xy$-plane's axes).
Thus, the second electron in NSI leaves the atom approximately half
laser cycle after the first one. The dynamics of the classical test
particles corresponds to the temporal and spatial evolution of the
quantum probability density as  described in Section \ref{qmresults}.

In Figure \ref{classhigh} we show a representative example for
double ionization at a higher
field strength ($\Edach=1.0$). At this field strength the
sequential pathway is more probable than NSI. The
temporal delay between the ionization of the two electrons is greater
(1--3 cycles) but still small since our pulse is ramped strongly over
3 cycles only. In longer pulses the temporal delay between the ejection of
the two electrons would be even greater.

\section{Conclusions} \label{concl}
In this paper we have confirmed the recently proposed mechanism for the
NSI process, namely the ionization of the second electron by
Coulomb-interacting  
with the outer partner. We have traced the  NSI scenario 
by observing the quantum mechanical probability density as it evolves
in space and time. We have identified the {\em Coulomb
repulsion assisted laser acceleration} of the inner electron 
to be responsible for NSI. Our classical simulations
have supported this point of view and contributed a detailed picture of
how NSI happens in terms of one particle energies and trajectories.
Moreover, we have shown that ``non-sequential'' means ``within half an
optical cycle'' and that NSI is not an essentially quantum mechanical effect. 
We showed that in the frequency and pulse duration regime under
consideration, SAE ionization yields are in poor agreement with the
exact results. 
TDDFT reproduces the total ionization probability very
well, even for the simplest effective potential available, namely the
Hartree-potential. TDDFT fails if one claims physical relevance for the
Kohn-Sham-orbitals by separately constructing single and double
ionization yields. However, since generally  only the knowledge of the total electron density
is necessary, e.g.\ for the calculation of high
harmonics generation,  TDDFT should produce very accurate results \cite{erhard}.

\section*{Acknowledgement}
The author would like to thank P.\ Mulser and R.\ Schneider for helpful
discussions. This work has been supported by the European Commission
through the TMR Network SILASI (Super Intense Laser Pulse-Solid
Interaction), No.~ERBFMRX-CT96-0043.

\begin{figure}[ht]
\caption{\label{potentials}}
{  The 2D potential $V(x,y)$ \reff{2dpot} for
$\Edach=0.0$ (a), $\Edach=0.1$ (b) and $\Edach=0.616$ (c). At $\Edach=0.1$ the initial ground
state level cuts the effective potential within the single ionization channels. For $\Edach=0.616$ the ground state level
even exceeds the potential ridge along $x=y$. At latest at that field
strength strong double ionization is expected.  }
\end{figure}

\begin{figure}[ht]
\caption{\label{groundstate}}
{  Contour plot of the ground state probability density. The energy is
$\energy_0=-2.897$. The electron-electron repulsion along $x=y$ clearly 
leaves its fingerprint on the wave function (butterfly-shape). Such an
asymmetric shape is absent in corresponding Hartree-Fock
\cite{pindzola_ii} or simple DFT ground states.}
\end{figure}

\begin{figure}[ht]
\caption{\label{timeseries}}
{  The probability density after $n$ optical cycles for
$\Edach=0.3$. One looks perpendicularly from above onto the
illuminated $xy$-plane and the logarithmically scaled probability
density $\vert \Psi(x,y,t)\vert^2$. Till $n=2$ mainly single ionization takes place
(probability density along the axes). Afterwards also regions $\vert x\vert,
\vert y\vert > 5$ are occupied which corresponds to double
ionization. The Coulomb-repulsion ridge along $y=x$ can be clearly
identified at later times. }
\end{figure}

\begin{figure}[ht]
\caption{\label{integrationareas}}
{  The areas to be integrated over in order to calculate the
probabilities for (from left to right) total, double and single
ionization are shaded. The parameter $a$ was chosen to be 5~a.u.}
\end{figure}

\begin{figure}[ht]
\caption{\label{timeseries_detail}}
{  The probability density for $\Edach=0.2$ at $n=2.5,2.75,3,3.25,3.5,3.75$
optical cycles. At time $n=3$ two probability density jets are
emitted into the region $x,y>0$. At $n=3.25$ they are partly flown
back and density appears in regions $x<0, y>0$ and $y<0, x>0$. }
\end{figure}

\begin{figure}[ht]
\caption{\label{timeseries_detail_0.7}}
{  The probability density for $\Edach=0.7$ at $n=1.5,1.75,2,2.25,2.5$
optical cycles. Double ionization is mainly due to the strongly tilted
$xy$-plane.}
\end{figure}

\begin{figure}[ht]
\caption{\label{ioni_yields_I}}
{  The 2D calculation results for total, single and double ionization
(bold, +). The SAE result for the He$^+\to$He$^{++}$ is also drawn
($\Diamond$). Multiplication with $P$ (see text) leads to the expected
sequential double ionization degree ($*$) as long as $P^{++}\ll P^+$. The $\Pseq$ and $
P\cdot\Pseq$ curves intersect the correct $P^{++}$ result since depletion
of the He$^+$ ions is not taken into account. }
\end{figure}

\begin{figure}[ht]
\caption{\label{ioni_yields_II}}
{  SAE results for the outer electron compared with the exact 2D yields
(bold, +). The curve plotted with connected $*$ was calculated using
$\Zeff=1.117$ and $\epsiloneff=\epsilon=0.55$. For the $\Diamond$-curve
$\Zeff=1$, $\epsiloneff=0.398$ was chosen. In both cases the SAE
ionization degrees are too high.} 
\end{figure}

\begin{figure}[ht]
\caption{\label{tddftcomp}}
{  Comparison of the TDDFT results with the exact 2D solutions. The total
ionization degree $\Pks$ ($\bigtriangleup$) matches nicely the exact curve
(bold, +). If one claims physical relevance of the Kohn-Sham orbitals
one gets the $\Diamond$-curves for $\Pkstot$ (total), $\Pks^+$
(single) and $\Pks^{++}$ (double)
ionization which agree poorly with the exact probabilities (see text
for a discussion).   }
\end{figure}

\begin{figure}[ht]
\caption{\label{dftprobab_i}}
{  The TDDFT probability density at the end of the pulse for
$\Edach=0.3$. The gridlike pattern is due to the construction as a
pure product of the Kohn-Sham orbitals. There is no displacement of
probability density along $y=x$, of course. }
\end{figure}

\begin{figure}[ht]
\caption{\label{highfieldprobabs}}
{  The probability densities at the end of the laser pulse for
$\Edach=0.7$. In plot (a) the exact density, in (b) the
TDDFT result is presented. The exact density (a) develops a gridlike
pattern since both electrons were ionized rapidly and subsequently
behaved like almost free electrons, i.e., the wave function becomes
more and more a pure product of single particle wave functions. 
However, the agreement with the product
of the Kohn-Sham-orbitals (b) is poor even at those high fieldstrengths.  }
\end{figure}

\begin{figure}[ht]
\caption{\label{classioni}}
{  The classical yields (bold, $*$) for $\Pcl$ (total), $\Pcl^+$
(single) and $\Pcl^{++}$ (double)
ionization. The single ionization is strongly overestimated (compare
with the exact quantum mechanical results, drawn dashed) due to the classical effect
discussed in the text. The classical SAE results for the outer
($\Ptildecl^+$, $\Box$) and the inner electron ($\Ptildecl^{++}$, $\bigtriangleup$) are also
plotted. There is classical NSI at $I=0.04$ even when no sequential single
ionization should occur! }
\end{figure}

\begin{figure}[ht]
\caption{\label{classsingle}}
{  A representative example for classical single ionization in terms
of the single particle energies (\ref{single_energy_x},\ref{single_energy_y}). The
inner electron (solid) drops below the quantum mechanical He$^+$ binding
energy $-1.920$. This leads to an overestimation of the single ionization
probability since the outer electron (dashed) can gain more energy during
Coulomb collisions  (see peaks in the energy-curves) with the inner
electron as allowed quantum mechanically. }
\end{figure}

\begin{figure}[ht]
\caption{\label{classicalnsi}}
{  Two representative classical NSI scenarios in terms of
the single particle energies 
(\ref{single_energy_x},\ref{single_energy_y}). 
 }
\end{figure}

\begin{figure}[ht]
\caption{\label{classicalnsi_traj}}
{ The particle trajectories corresponding to Figure
\ref{classicalnsi}. The electrons become free within approximately $\halb$ an
optical cycle.
Before both electrons
ionize they move together in the same direction for a quarter of a
cycle until one electron turns and finally vanishes in the opposite
direction. This has to be compared with the quantum dynamics in Figure
\ref{timeseries_detail}. }
\end{figure}

\begin{figure}[ht]
\caption{\label{classhigh}}
{  A representative example of classical sequential double ionization at
$\Edach=1.0$. 
The temporal delay between the ejection of the two electrons is 3 half
cycles and would be even greater in a more adiabatically ramped pulse.}
\end{figure}

\end{document}